\pdfoutput=1 
\documentclass{JINST}
\let\ifpdf\relax
\usepackage[numbers]{natbib}
\usepackage{amsmath}
\DeclareGraphicsExtensions{.pdf,.png,.jpg}

\title{The brighter-fatter effect and pixel correlations in CCD sensors}

\author{P. Antilogus$^a$,
P. Astier$^a$\thanks{Corresponding author.}, P. Doherty$^b$, A. Guyonnet$^a$~
and N. Regnault$^a$\\
\llap{$^a$}LPNHE/IN2P3/CNRS,\\
  UPMC, 4 place Jussieu F75005 Paris, France\\
\llap{$^b$}Department of Physics, Harvard University, \\
          17 Oxford Street, Cambridge, MA 02138, USA \\
E-mail: \email{pierre.astier@in2p3.fr}}

\abstract{We present evidence that spots imaged using
astronomical CCDs do not exactly scale with flux: bright spots tend to
be broader than faint ones, using the same illumination pattern. We
measure that the linear size of spots or stars, of typical size 3 to 4
pixels FWHM, increase linearly with their flux by up to 2 \% over the
full CCD dynamic range. This brighter-fatter effect affects both
deep-depleted and thinned CCD sensors. We propose that this effect is
a direct consequence of the distortions of the drift electric field
sourced by charges accumulated within the CCD during the exposure and
experienced by forthcoming light-induced charges in the same exposure.
The pixel boundaries then become slightly dynamical: overfilled pixels
become increasingly smaller than their neighbors, so that bright star
sizes, measured in number of pixels, appear larger than those of faint
stars. This interpretation of the brighter-fatter effect implies that
pixels in flat-fields should exhibit statistical correlations, sourced
by Poisson fluctuations, that we indeed directly detect. We propose to
use the measured correlations in flat-fields to derive how pixel
boundaries shift under the influence of a given charge pattern, which
allows us in turn to predict how star shapes evolve with flux. We show
that, within the precision of our tests, we are able to quantitatively
relate the correlations of flat-field pixels and the broadening of
stars with flux. This physical model of the brighter-fatter effect
also explains the commonly observed phenomenon that the spatial variance
of CCD flat-fields increases less rapidly than their average.}


\begin{document}

\section{Introduction}
\label{sec:intro}
The astronomical use of CCDs very commonly assumes that 
these sensors slice the incoming light along a regular 
lattice of pixels, each of which then responds independently
of what adjacent pixels receive. Non-linearity usually
refers to the reponse of the electronic chain, which
can be (and often is) corrected independently for each pixel.

In this representation, the images scale precisely with the incoming flux
or staring time on a static scene, possibly after some linearity
restoration is applied at the pixel level. We will discuss in this
contribution a tiny violation of this idealistic picture of CCDs: images of
stars (or laboratory spots) do not exactly scale with the incoming
flux, even with a prefectly linear readout chain.  With increasing
flux, images of spots or stars tend to broaden: the measured sizes
grow linearly with flux. Since the effect is almost isotropic (though not
exactly), there is a strong inclination to attribute it to
sensors rather than to the read out chain. All attempts to detect the
effect we know about have been successful, and the effect does not
seem to only affect deep-depleted CCDs.

There is another known departure from the independent pixels
representation: the variance of a flat-field image does not grow linearly with the 
exposure average (as Poisson statistics robustly predict), but the 
variance ``flattens out'', as if some mechanism washing the contrast
were at play. This phenomenon is sufficiently similar to the brighter-fatter
effect to attempt to bridge these two effects. 

The flattening of the photon transfer curve (PTC i.e. the spatial
variance of flat-fields as a function of their average) is known to be
reduced or to almost disappear by summing neighboring pixels into larger
pixels (e.g. \cite{Downing06}). This indicates that the variance of a
sum of neighboring pixels is larger than the sum of their variances,
and adjacent pixels should therefore have (positive) covariances. 
We detect these covariances. They contain more information
than the flattening of the PTC, and we will hence study 
pixel correlations in flat-field images rather than the 
``missing variance'' one can extract from the shape of the PTC. 

Large scale imaging surveys, underway and anticipated, vitally require
precise knowledge of their Point Spread Function (PSF). Regarding the
accuracy of the PSF shape and size, the most demanding application is
arguably the measurement of the shear of background galaxy images induced
by mass density gradients along the line of sight, in order to
constrain these mass gradients, and in turn cosmology. Usually, the
design of the data reduction chain assumes that stars collected in the
science image are faithful models of the PSF, so that the effects of
the observing system (possibly including atmosphere) on objects shapes
can be measured and ``factored out'' using these stars. We will argue
that stars are acceptable PSF templates only for applications that can
accommodate an inaccuracy of the PSF size at the percent scale. In
particular, studying the correlations of the shear field measured from
galaxy shapes now requires a PSF typically understood at the $10^{-3}$
level or better (see e.g. the contribution of M. Jarvis at this
workshop\footnote{The presentations at this workshop can be found on 
\href{http://www.bnl.gov/cosmo2013/}{http://www.bnl.gov/cosmo2013/}}). 
The requirements on PSF ellipticity accuracy are as low as
$2\ 10^{-4}$ for the Euclid space mission
\cite{EuclidRB}.

We attempt in this contribution to bridge the brighter-fatter effect
and the correlations in flat-fields, assuming that both are a
consequence of alterations of drift lines in the CCD bulk induced by
charges accumulating in the potential wells of the sensor during
the exposure. These drift field alterations can be regarded as
dynamical modifications of pixel boundaries\footnote{The
``tree-rings'' phenomenon (see the contributions of e.g. G. Bernstein,
E. Magnier \& R. Lupton at this workshop) is probably due to a similar
phenomenon, but induced by a static distortion of the drift field
sourced by spatial and static variations of the silicon bulk
resistivity.}.  We wish to set up this bridge in order to derive the
details of the brighter-fatter effect from pixel correlations in
flat-fields. This would allow one to account for the brighter-fatter
effect in the data reduction procedures, for a variety of image
qualities and PSF shapes, and beyond the growth of second moments.

In this paper we proceed as follows. First we expose the characteristics
of the brighter-fatter effect (\S~\ref{sec:detection}). We then discuss
statistical pixel correlations in flat-fields (\S~\ref{sec:correlations}).
We propose in \S~\ref{sec:calculations} that Coulomb forces
sourced by already collected charges on drifting ones are the
source of both effects, and propose a parametrized model 
that explains both. In \S~\ref{sec:from-flat-to-bf},
we derive the model parameters from flat-field images and then predict
the brighter-fatter slope for three different sensor types.
We summarize and discuss possible further work in \S~\ref{sec:discussion}.

\section{Detecting the brighter-fatter effect}\label{sec:detection}

\begin{figure}[tbp]

\centering
\includegraphics[width=.8\textwidth]{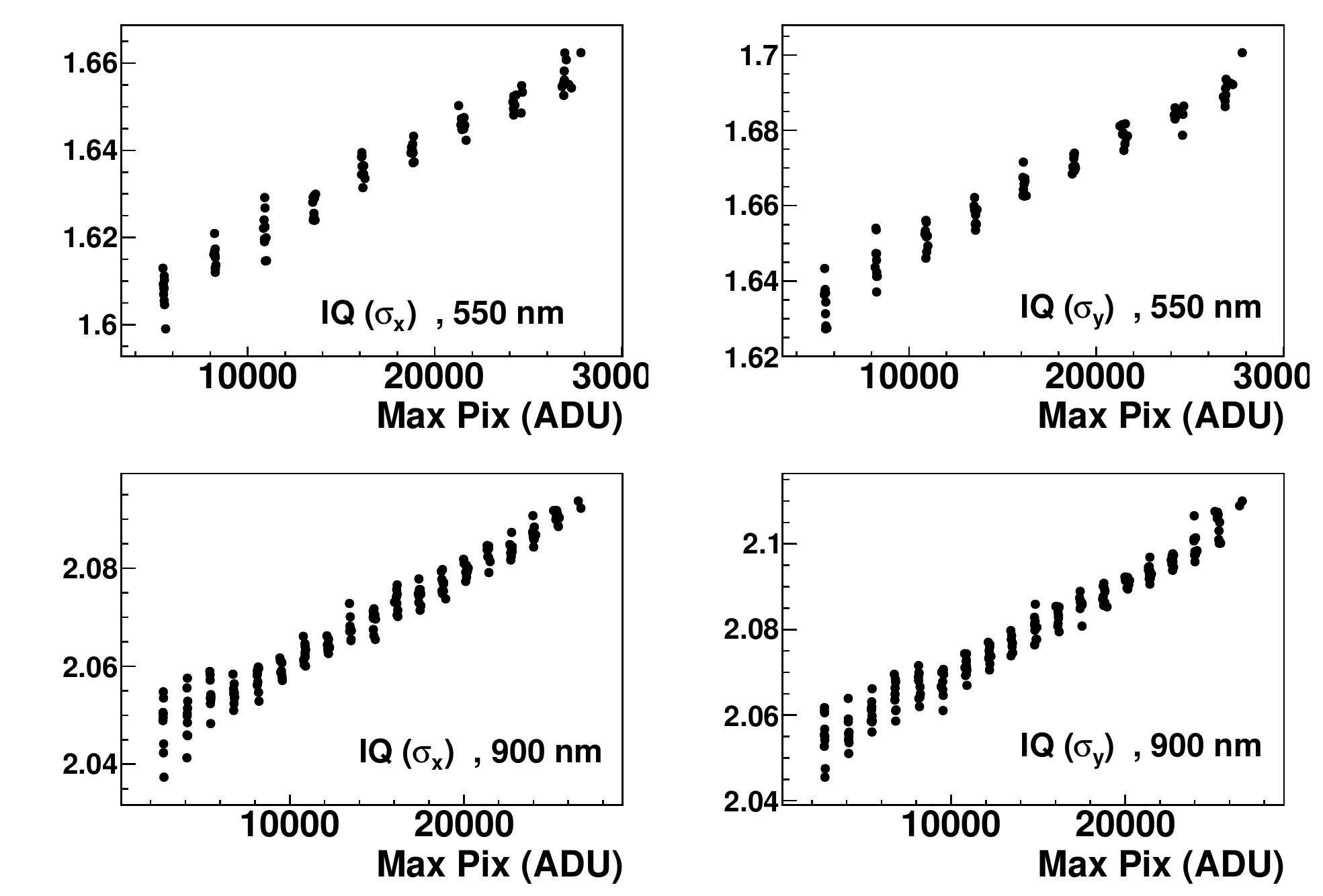}
\caption{Measured sizes (pixels rms) of a laboratory spot along x and
  y coordinates of the E2V CCD250, as a function of the spot peak
  flux in ADUs (the gain is about 5 e-/DN). The relative increase of size is
  more than 2\% over the full dynamic range, and slightly larger along
  $y$ direction (parallel transfers) than along $x$ (serial transfers). The
  relative size increases are similar at two different
  wavelengths. Diffraction in the illumination system is likely
  causing the difference in size of the spots at these two wavelengths.
\label{fig:bf3}}
\end{figure}

It is in fact fairly easy to detect the ``brighter-fatter'' effect.
One either needs a set of images of a spot of increasing integrated
flux (in the lab, that can be done by varying the open shutter time),
or an astronomical image with the full range of star brightnesses. One
then measures the second moments of spots or stars using an estimator
whose bias does not vary with S/N (e.g. eq. 1
from \cite{Astier13}). One can then observe that the linear sizes of
spots increase linearly with flux, in both directions.  We have
detected the effect in the laboratory using spots projected 
on an E2V CCD250, a candidate sensor for the Large Synoptic Survey 
Telescope (LSST) camera. Our measurements,
displayed in fig. \ref{fig:bf3} show that the size of the spots
rises linearly with their peak flux, in both directions. The size increase
amounts to 2 to 3 \% at 150 ke. We have also
detected the effect in on-sky data from the thinned sensors of Megacam on
CFHT (type 42-90 from E2V), where it amounts to 0.5\% over the full dynamic range
(saturation occurs at 100 ke) and in data from
DECam on the CTIO Blanco Telescope (using the publicly available science verification
data), where it amounts to 1.5\%. DECam is populated with CCDs manufactured by LBNL, 
which are deep-depleted 250~$\mu$m thick sensors. We hence measure variations 
of size
in the \% range, and consistently slightly steeper (typically 20 \%)
along $y$ (parallel shifts), than along $x$ (serial shifts).  At this
workshop, the effect was also reported by R. Lupton about the
Hamamatsu sensors S10892-02 populating the Hyper Suprime Cam focal plane. 
We do not know of any sensor that does not exhibit the effect.  For images 
from the sky, one must take care to isolate the ``color independent''
brighter-fatter effect from the more trivial spectral dependence of
the PSF coupled to statistical relations between color and brightness
in the star sample used for the measurement. We display our
measurements in fig \ref{fig:bf3}.

One can think of at least two applications that require knowledge of the
PSF size at the \% level or better: correlations of the gravitational
shear over large areas on the sky, and accurate PSF photometry, in
particular when measuring supernova light curves for estimating their
distance. The brighter-fatter effect is a manifestation of a more
general departure of star images from the actual PSF, and some
handling of the effect beyond the size of the PSF should be devised.
Ground-based applications face one extra complication : the slope of
the effect depends of the image quality itself, which potentially
changes at every exposure. A physical understanding of the origin of
the effect would then allow one to extract an unbiased PSF
model from stars observed in virtually all observing conditions.

\section{PTC non-linearity and statistical correlations in flat-fields}
\label{sec:correlations}
\begin{figure}[tbp] 
\centering
\includegraphics[width=.8\textwidth]{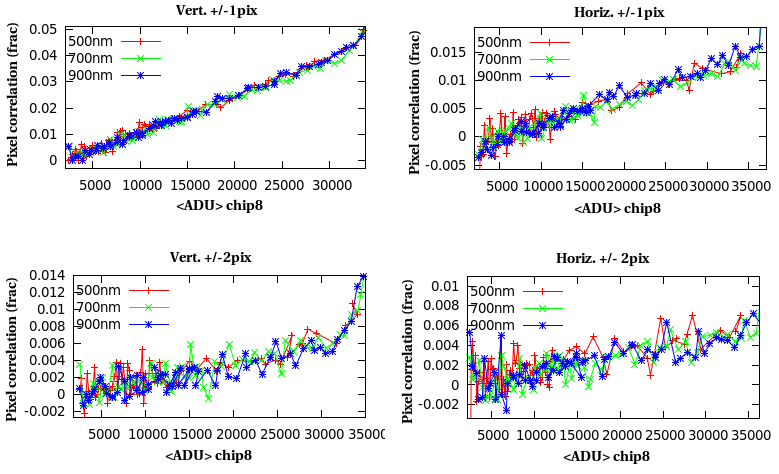}
\caption{Correlation coefficients of flat-field pixel pairs as a 
function of illumination of the flat-field, at three different wavelengths.
These correlations rise linearly with illumination and are achromatic.
They decay with distance, and are larger along y (CCD columns) than along
x, especially for the nearest neighbors. These correlations are indeed
measured on differences of flat-field exposures with the same average.
These measurements were carried out on the E2V CCD250 sensor, candidate for LSST.
\label{fig:quatre-correlations}}
\end{figure}

There exists one other known departure from the picture of independent
pixels: the spatial variance of flat-fields is not exactly
proportional to the average illumination in the flat-field, but
``flattens out'': high-illumination flat-fields exhibit a lower
variance than would be expected from extrapolating measured variances at lower
illumination. This apparent violation of Poisson statistics tends to
vanish when one groups neighboring pixels into larger pixels
\cite{Downing06}. This strongly suggests that neighboring pixels of
flat-fields have statistical correlations, which we indeed found.  As
shown in figure \ref{fig:quatre-correlations}, the correlation
coefficients between nearby pixels appear to be a linear function of
illumination, up to saturation, and decay rapidly with
separation. They look roughly isotropic, except for the nearest
neighbors where the vertical correlations seems constantly larger than
the horizontal correlation (as shown as well by R. Lupton for HSC at
this workshop). These correlations also look mostly achromatic.

The fact that the correlation coefficients rise linearly with
the flat-field average implies that covariances rise quadratically
with illumination and hence that the PTC  should exhibit
a (small) quadratic correction to Poisson statistics. For the E2V CCD250 
we find that the size of the measured covariances fully explains 
the ``missing variance'' on the PTC.

\begin{figure}[tbp] 
\centering
\includegraphics[width=.8\textwidth]{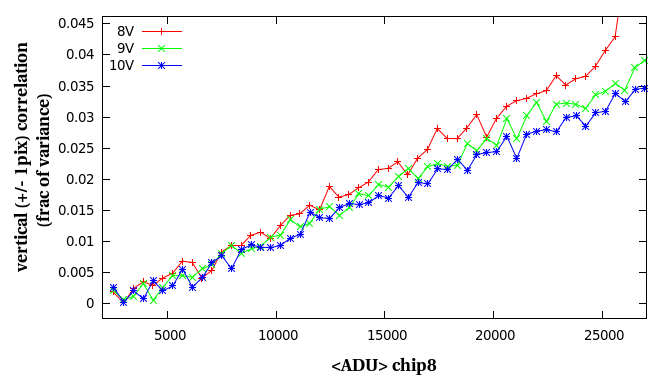}
\caption{Same plot as figure \protect\ref{fig:quatre-correlations}, at 700nm, but
for different clock voltages during collection. This indicates
that correlations are likely related to electrostatics. 
\label{fig:correlations-8-9-10}}
\end{figure}

The correlations depend on electrostatics in the CCD: lowering
the clock voltages applied during collection increases the correlations
in the vertical ($y$) direction (fig. \ref{fig:correlations-8-9-10}).
This fact, along with an almost isotropic correlation pattern beyond
first neighbors, strongly suggests that the origin of
the correlations lies in the sensor itself.

\section{Coulomb forces in CCDs}
\label{sec:calculations}.

As already presented, we argue that forces produced by stored charges
and experienced by subsequently collected charges are the common cause of both
the brighter-fatter effect and the pixel correlations in flat-fields. We
first present numerical electrostatic calculations from first
principles, that show that the observed size of both effects is
compatible with reasonable values of the key parameters. The reach of
these simulations is however limited by our restricted knowledge of
several parameters that determine the electrostatics of CCDs. We hence
develop a parametrized model of drift field alterations that aims at
describing both phenomena using the same algebra.

\subsection{Simplified calculations from first principles}
\label{sec:simu}
\begin{figure}[tbp] 
\centering
\includegraphics[width=.9\textwidth]{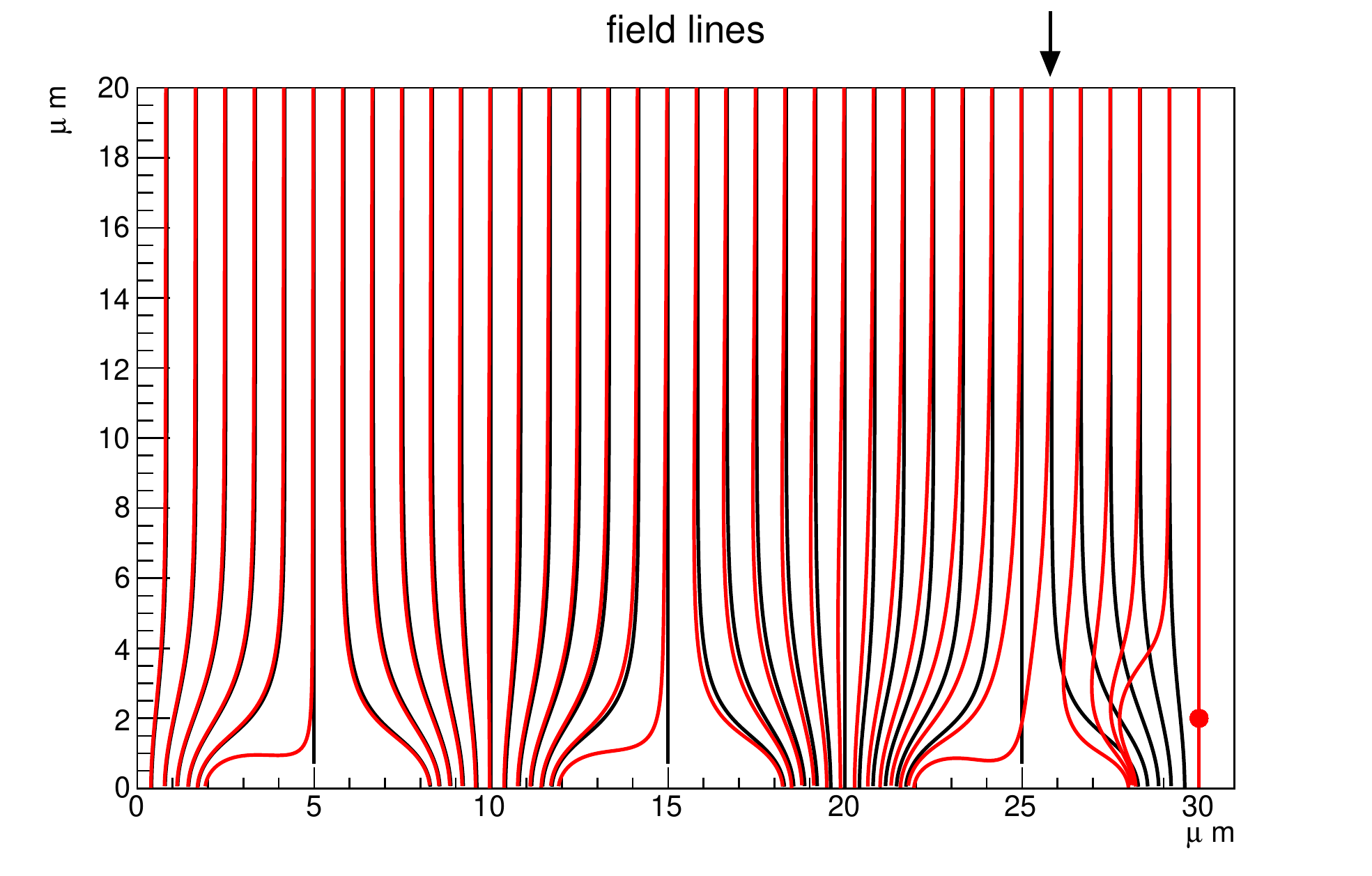}
\caption {Computation of field lines for the E2V CCD 250 geometry,
  with (red) and without (black) a 50~ke charge positionned at the
  bottom right (red spot). An electron drifting along the arrow drawn
  at the top of the picture goes either left or right depending on the
  presence of the charge, which illustrates that the rightmost
  pixel shrinks when storing more charge than its neighbor. The stored charge
  also shifts farther pixel boundaries.  Note that we have only drawn
  the CCD collection area, and the total device thickness is
  100~$\mu$m rather than the bottom 20~$\mu$m drawn on the vertical axis.
  \label{fig:field-lines}}
\end{figure}

We have conducted a simplistic 3D electrostatic simulation of a CCD in
order to evaluate drift lines and how they are altered when charges are
present in the CCD. Our aim is not to reproduce in all details the
measurements, but to assess if the size and other characteristics of
the observed effects (both the brighter-fatter effect the pixel
correlations) can be reproduced by the same calculation. We have
simulated the E2V CCD250 geometry, i.e. 100 $\mu$m thick square
pixels 10 $\mu$m on a side, in n-type silicon. We have approximated
the intrinsic silicon as free of charges, and we have implemented the
split across columns using the same potential pattern than the one
that defines rows: our simulation clearly does not aim for a high
fidelity.  Given the applied voltages, we have then solved numerically
for the potential, and computed drift trajectories. In particular, we
have located the pixel boundaries by tracking positive charges
backwards from pixel separations on the charge collecting face.  Using
the image series technique (see e.g. \cite{Durand}), we have computed
the potential of a charge between equipotential planes and added it to
the potential created by the voltages applied to the sensor. We can
then evaluate how drift lines are altered when some charge is stored 
in the CCD. We illustrate in figure \ref{fig:field-lines} how some charge
added in the CCD alters drift lines.  From these alterations, we can
derive both the brighter-fatter effect and pixel correlations in
flat-fields.

\begin{figure}[tbp] 
\centering
\begin{minipage}{0.5\textwidth}
\includegraphics[width=.9\textwidth]{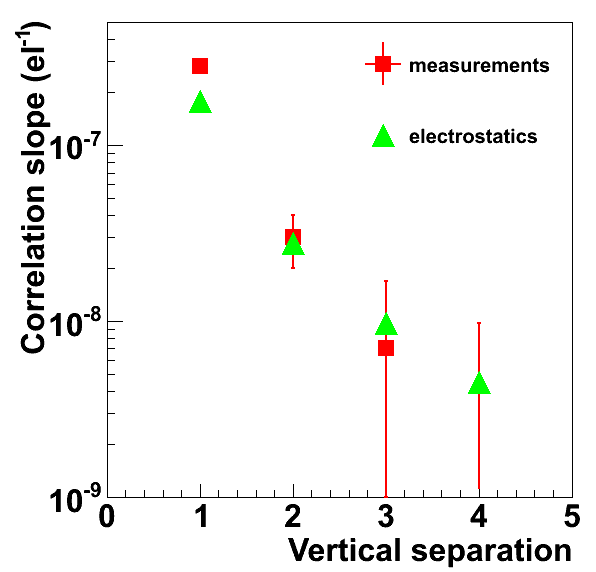}
\end{minipage}\begin{minipage}{0.5\textwidth}
\includegraphics[width=.9\textwidth]{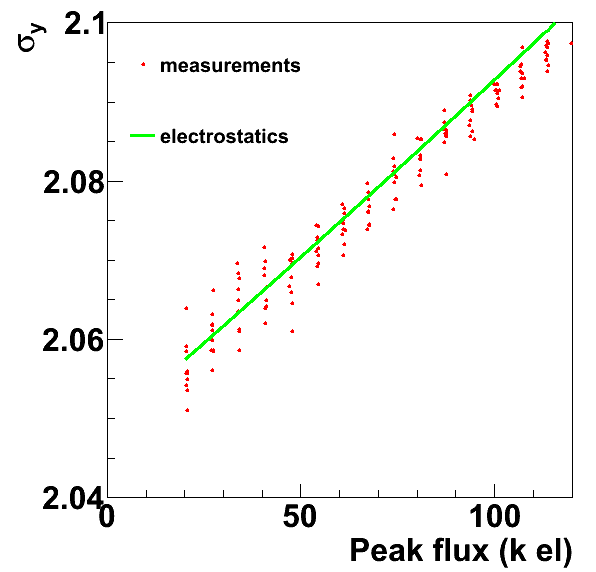}
\end{minipage}
\caption{Comparison of our crude electrostatic simulation and
measurements for the correlations along y (left) and the brighter-fatter
slope (right). This indicates that Coulomb forces in CCDs have 
the right scale for causing both effects. 
\label{fig:comp-sim-data}}
\end{figure}

Our crude simulation only has one adjustable parameter: the distance
between the gate plane (which has imposed potentials) and the charge
stored in a pixel. This distance is often referred to as the depth of
the buried channel. We have found a fair match between our simulations
and the data from the E2V CCD250 for a distance of 2.5
$\mu$m. The comparison of anticipated and measured effects
is displayed in figure \ref{fig:comp-sim-data}.
The fact that the same simulated setup can reproduce the observed slope
of the brighter-fatter effect {\it and} the observed scale of
correlations constitutes an encouraging indication that Coulomb forces
are the possible common and dominant cause of both effects.
 
From these simulations, we were able to evaluate that the alterations
of drift trajectories mostly happen in the last microns above the
clock stripes (see fig. \ref{fig:field-lines}), for a variety of geometries 
and voltage values. Because this very last part of the electron drift is experienced 
by all electrons, irrespective of the photon conversion depth, this
Coulomb force ansatz naturally explains the observed achromaticity of
correlations.

Note that when comparing electrostatic simulations to real data, one
has to account for the fact that real images depict the final state
of the charge distribution in the CCD, but the collection of electrons
happened under less contrasted conditions, because less charge was
residing at the bottom of the CCD during most of the exposure than at 
its end. It is fairly easy to account for this effect: on average, the charge
contrasts during the exposure are half of what they are at the end,
and to first order, the perturbations to pixel boundaries are
proportional to the charge contrasts. So, when computing perturbations
from an electrostatic simulation, one just has to halve the effects
computed for the image ``final state''.

Making the electrostatic simulation of CCDs more and more realistic
might seem a possible avenue to account for the brighter-fatter
effect. However, the outcome of the fabrication process is not
necessarily known precisely enough, and CCD vendors might regard some
key design features as proprietary. We could then consider using the
correlations in flat-fields to predict or at least constrain the
brighter-fatter effect. We have then developed a generic model in
order to describe both effects using the same algebra.

\subsection{A generic model for electrostatic distortions of drift fields}
\label{sec:model}
In this section, we approximate the electrostatic influence of charges
in the CCD as pixel boundary shifts. We model the displacement of
the effective boundaries of a pixel, labeled $(0,0)$, caused by a charge
$q_{i,j}$ in the potential well of a pixel at position $(i,j)$ as:
\begin{equation}
\delta^X_{i,j} = q_{i,j}  a^X_{i,j}
\end{equation}
where we have expressed that the (perturbating) electric field 
due to a charge is proportional to this charge (which is exact), 
and we have approximated alterations to drift trajectories as first order
perturbations. 
$X$ indexes the four boundaries of the pixel $(0,0)$,
and we label each boundary by the coordinates of the pixel
that shares it with (0,0): $X \in \{(0,1),(1,0),(0,-1),(-1,0)\}$.
The $a^X_{i,j}$ coefficients satisfy symmetries:
\begin{align}
a^X_{i,j} &= a^{-X}_{-i,-j}   & \text{(parity)}\\
a^{0,1}_{i,j} &= -a^{0,-1}_{i,j-1} & \text{(translation invariance)}
\end{align}
The  $a^X_{i,j}$ also obey a sum rule:
\begin{equation}
\sum_{i,j} a^X_{i,j} = 0 , \forall X \label{eq:sum-rule}
\end{equation} 
because the electric field induced on a pixel boundary (and hence the
displacement of the latter) is null if all charges $q_{i,j}$ are
equal. Each boundary of the pixel $(0,0)$ shifts under the influence
of all charges. Because electric fields add up, the displacement reads:
\begin{equation}
\delta^X  = \sum_{i,j} \delta^X_{i,j} =  \sum_{i,j} q_{i,j}  a^X_{i,j} \nonumber
\end{equation}

We call charge transfer 
the difference between charge contents with and without the
perturbating electric fields. 
The displacement of the pixel boundary $\delta^X$ induces a charge transfer
between the pixel $(0,0)$ and the pixel $X$, where $X \in$ 
$\{(0,1),(1,0),(0,-1),(-1,0)\}$. To first order, this charge
transfer is proportional to both the pixel boundary displacement
and to the charge density flowing on this boundary. For a well sampled image
(i.e. the charge distribution impinging on a pixel does not vary rapidly
within this pixel), we can approximate the charge density drifting
on the boundary between pixel $(0,0)$ and its neighbor $X$ 
as:
\begin{equation}
\rho_{00}^X \propto (Q_{0,0}+Q_X)/2 \nonumber
\end{equation}
so that the net charge transfer due to perturbating electric fields
between pixel $(0,0)$ and its neighbor $X$  reads:
\begin{equation}
\delta Q^X_{0,0} = \sum_{i,j} a^X_{i,j} Q_{i,j} (Q_{0,0} + Q_X) \label{eq:delta-q-x}
\end{equation}
The expression is non-linear with respect to the charge distribution:
the charges $Q_{i,j}$ are the source charges, and the expression
$(Q_{0,0} + Q_X)$ approximates the test charges. We have deliberately
incorporated all proportionality coefficients into the $a^X_{i,j}$
factors. The perturbed charge in pixel $(0,0)$ reads:
\begin{equation}
Q'_{0,0} = Q_{0,0} +\delta Q_{0,0}
\end{equation}
with 
\begin{equation}
\delta Q_{0,0} = \sum_X \delta Q^X_{0,0} =  \sum_X \sum_{i,j} a^X_{i,j} Q_{i,j} (Q_{0,0} + Q_X)
\label{eq:full-model}
\end{equation}
where $X$ runs over the four boundaries of pixel $(0,0)$, i.e.  $X \in \{(0,1),(1,0),(0,-1),(-1,0)\}$

In order to evaluate the statistical correlations in flat-field
exposures introduced
by the charge-induced perturbations of drift trajectories, we wish
to evaluate 
\begin{equation}
Cov(Q'_{i,j}, Q'_{0,0}) = Cov(Q_{i,j}, \delta Q_{0,0}) + [(i,j)\leftrightarrow(0,0)] + O(a^2) \label{eq:cov}
\end{equation}
where $O(a^2)$ stands for expressions quadratic in the $a^X_{i,j}$
coefficients of expression \ref{eq:full-model}. We will stick to
first order perturbation expressions, as real data indicates that
this is justified. In the case where the pixel $(i,j)$ is not a nearest 
neighbor of $(0,0)$, we have:
\begin{align}
Cov(Q_{i,j}, \delta Q_{0,0}) &= Cov \left( Q_{i,j}\ ,\ \sum_{k,l} a^X_{k,l} Q_{k,l} (Q_{0,0} + Q_X) \right) \nonumber \\
 &= Var[Q_{i,j}] \sum_X a^X_{i,j} E[Q_X+Q_{0,0}] \label{eq:delta-q-x2}
\end{align}
where $E$ stands for the expectation value. 
For a flat-field illumination of average content
$\mu$ and variance $V$, the above expression reads:
\begin{equation}
Cov(Q_{i,j}, \delta Q_{0,0}) = 2 \mu V  \sum_X a^X_{i,j} \label{eq:half-cov1}
\end{equation}
In case the pixel $(i,j)$ is a nearest neighbor of pixel $(0,0)$, say $Y$, we 
have two terms in the covariance:
\begin{align}
Cov(Q_Y, \delta Q_{0,0}) &= Cov \left( Q_Y, \sum_X \sum_{i,j} a^X_{i,j} Q_{i,j} (Q_{0,0} + Q_X) \right) \nonumber\\
\begin{split}
  &=  Var[Q_Y]  \left[ \sum_{X \neq Y} a^X_YE[Q_{0,0} + Q_X]+\sum_{(i,j) \neq Y} a^Y_{i,j} E[Q_{ij}] \right] \\
    &\quad \hspace{1cm} + a^Y_Y Cov(Q_Y, Q_Y (Q_{0,0}+ Q_Y)) \label{eq:3-terms}
\end{split}
\end{align}
We have:
\begin{align}
 Cov(Q_Y, Q_Y (Q_{0,0}+ Q_Y)) &= \mu Cov(Q_Y,Q_Y)+ Cov(Q_Y,Q_Y^2) \nonumber \\
 &= \mu V + Cov(Q_Y,Q_Y^2)  \nonumber
\end{align}
If $Q_Y$ follows a probablity density function symmetric around its mean, we have
\begin{equation}
Cov(Q_Y,Q_Y^2)=2 \mu V \nonumber
\end{equation}
For a Poisson distribution, this becomes $Cov(Q_Y,Q_Y^2)=2\mu^2+\mu$ where the second
term is the skewness contribution, negligible for actual flat-fields. 

In the context of a flat-field, the three terms of \ref{eq:3-terms} become:
\begin{equation}
Cov(Q_Y, \delta Q_{0,0}) = \mu V \left[2 \sum_{X \neq Y} a^X_Y + \sum_{(i,j) \neq Y} a^Y_{i,j}\right] + 3 a^Y_Y \mu V
\end{equation}
Because of the sum rule (eq. \ref{eq:sum-rule}), the second term in $[\ ]$ reads $-a^Y_Y$
so that for neareast neighbors, we also have:  
\begin{equation}
Cov(Q_Y, \delta Q_{0,0}) = 2 V \mu  \sum_X a^X_Y \label{eq:half-cov2}
\end{equation}

 So whether or not $(i,j)$ is a nearest neighbor, we have for
the covariance between pixels in a uniform exposure of average $\mu$ and variance V:
\begin{equation}
Cov(Q'_{i,j}, Q'_{0,0}) = 4 V \mu  \sum_X a^X_{i,j} \label{eq:covar}
\end{equation}
to first order of perturbations. The factor of 4 compared with 
the factors of 2 in eq. \ref{eq:half-cov1} and \ref{eq:half-cov2} comes from the
symetric (second) term of eq. \ref{eq:cov}.

It turns out that in order to apply
this equation to $(i,j)=(0,0)$, we just have to account for the input ``variance'':
\begin{equation}
Cov(Q'_{0,0}, Q'_{0,0}) = V + 4 V \mu  \sum_X a^X_{0,0}
\end{equation}
where in practice, the $a^X_{0,0}$ are negative and are a plausible
cause of the photon transfer curve non-linearity.

So, electrostatic influence from collected charge induces covariances
between pixels in spatially uniform exposures that scale with both the average
and the variance of pixel contents. If one measures correlation
coefficients (ratio of covariance to variance), those are expected to
scale with the illumination level of the uniform exposure. This is precisely
what we observe in figure \ref{fig:quatre-correlations}.

With a realistic set of $a^X_{i,j}$ coefficients (e.g. from the
simulation \ref{sec:simu}) we observe that the
brighter-fatter effect is indeed linear with flux to an excellent
approximation (fig \ref{fig:comp-sim-data}).

\section{Closing the loop: deriving the brighter-fatter slope from 
flat-field correlations}
\label{sec:from-flat-to-bf}
\subsection{Solving for the model coefficients from flat-fields}

The first order model described in \S~\ref{sec:model}, is entirely
specified by the $a^X_{i,j}$ coefficients. Given their symmetries, we only
need those for $i,j \geqslant 0$. Furthermore, since these coefficients
describe Coulomb forces, they decay with distance, and we can 
neglect those describing large distance interactions. If we restrict
ourselves to $0 \leqslant i,j \leqslant n$, we have a priori $4(n+1)^2$ 
coefficients to determine. However, each boundary is shared by two pixels
and charge conservation reduces the number of coefficients by roughly
a factor of 2, more precisely to $(n+1)(2n+4)$. From the flat-fields,
we only have $(n+1)^2$ correlations to measure, which is short by about
a factor of 2.

In order to solve anyway, we have imposed ratios of coefficients
addressing similar source-test separations, using some reasonable
approximation for the dependence of electric fields on distance. Once
we have solved, we can evaluate the brighter-fatter slope by scaling
up faint spots or stars, and ``scrambling'' them using eq.
\ref{eq:full-model}. We have checked that the resulting
brighter-fatter slope is reasonably independent of the chosen
parametrization of distance dependence of the electric field.

\subsection{Comparing predicted and measured brighter-fatter slopes}

\begin{figure}[tbp] 
\centering
\includegraphics[width=.9\textwidth]{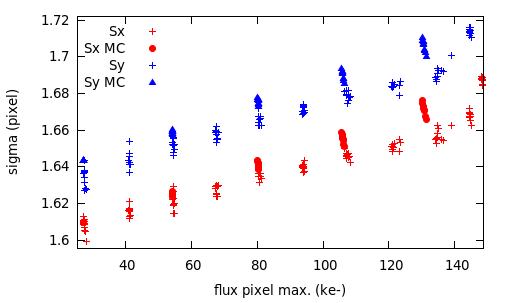}
\caption{Comparison of sizes of real spots (Sx, Sy) with
sizes of scrambled spots (Sx MC, Sy MC), as a function of their
peak flux. ``MC'' stands for 
a Monte Carlo propagation of shot noise affecting correlation
measurements in flat-fields. The predicted brighter-fatter
slopes are larger than the measured ones for both x and y directions,
for this data sample collected at 550 nm. The 900 nm data set 
displays very similar trends.
\label{fig:comp-bf-lsst}}
\end{figure}
\begin{figure}
\centering
\includegraphics[width=.9\textwidth]{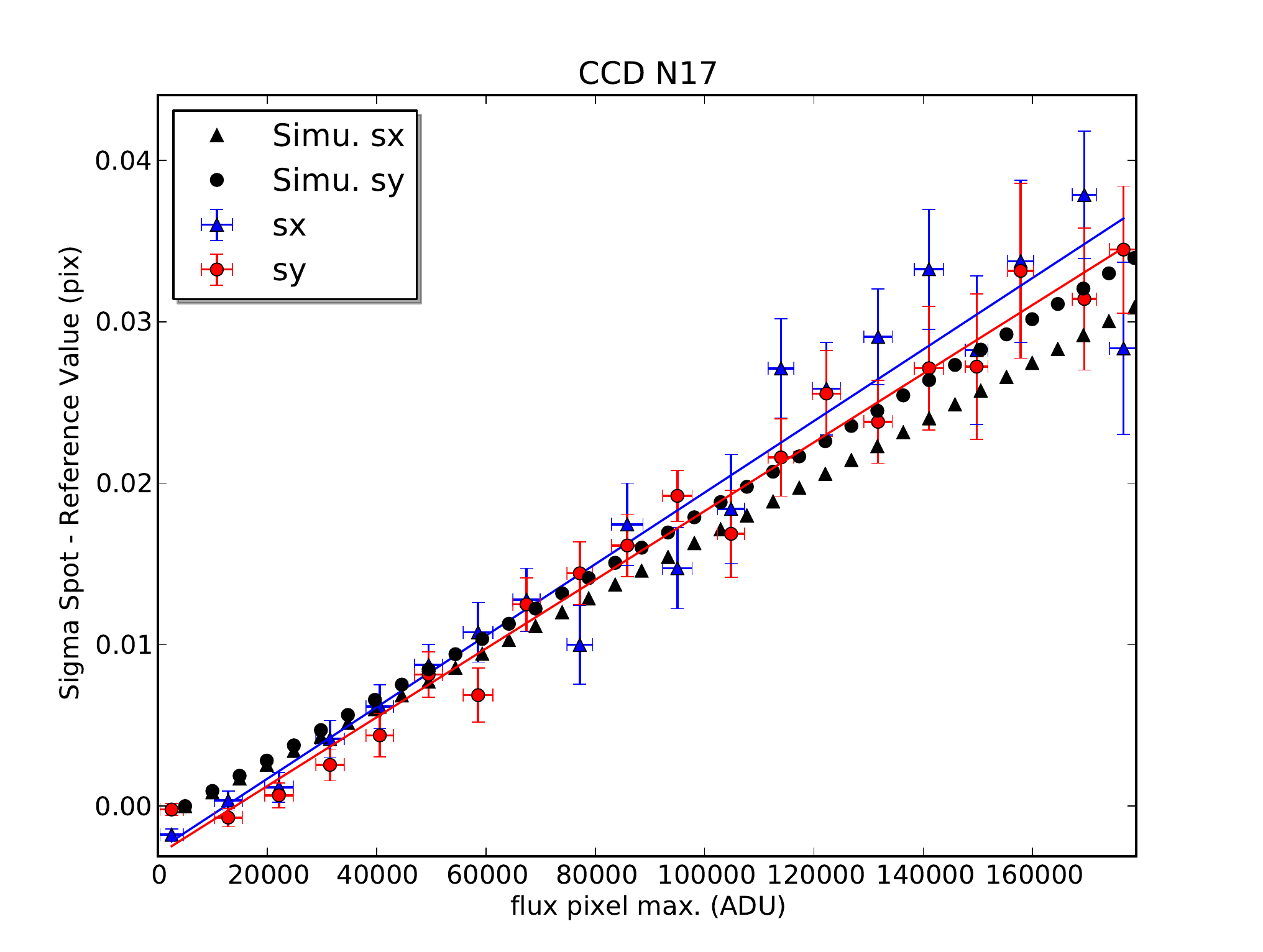}

\caption{
Comparison of sizes of real stars, and the evolution of the
scrambled PSF model (modeled as independent of flux) for CCD 17 of DECam.
We have subtracted the image quality of the 
image from each star measurement, which is on
average $\sigma_{IQ} \simeq 1.7$  for this data sample,
i.e the image quality is $\sim$4 pixels FWHM.
The agreement is fair, but statistically limited by the number of stars
we measured. When averaging over CCDs (see text), the slopes
disagree by $\sim$20\%.
\label{fig:comp-bf-decam}}
\end{figure}

The comparison between the brighter-fatter slopes predicted from 
flat-fields and the ones measured directly is done using the same data
used for the detection of the brighter-fatter effect. As just mentioned,
we transform faint spots or stars into bright ones by scaling them up
and transforming the image using eq.~\ref{eq:full-model}. In order to account
for the small amount of broadening already present in faint spots or stars,
we first apply the reverse transformation of eq.~\ref{eq:full-model}
to faint spots, which we approximate by the same expression with
opposite $a$ coefficients. We have checked that for the typical
size of the $a$ coefficients, flipping signs is a fair approximation
of the inverse transformation\footnote{Typically, a star with 100 ke
at peak that sees it size increase by $\sim$2\%, is decreased by the 
flipped transformation to its original size at the per mil level. 
So, for bright stars, applying successively the regular and
flipped transformations causes size changes that differ by
$\sim$5\%.}, 
especially for faint stars.

For the E2V CCD250, we compare the measured sizes of real spots and
scrambled spots in fig. \ref{fig:comp-bf-lsst}: the brighter-fatter
slopes are similar at the $\sim$20\% level, where only a fraction of
the difference can be attributed to the shot noise affecting the
measurements of correlations.  Indeed, the trend is very similar on
the independent data set taken at 900 nm.

We display a similar comparison for a selected DECam CCD in
fig. \ref{fig:comp-bf-decam}. The disagreement of slopes is of about
the same size, but exhibits the opposite sign, and might be affected
by the overall non-linearity of response (that we briefly discuss 
in \S~\ref{sec:discussion}). The precision is
limited by statistics of the star measurements. We have measured the
brighter-fatter slopes for all chips of the mosaic and find an rms of the
distribution about two times larger than expected from propagating the
shape measurement noise. We hence do not conclude yet that 
brighter-fatter slopes are compatible on different chips of the mosaic. 
We can however compare the average measured brighter-fatter
slope to the average one predicted from correlations, and find
the prediction from flat-fields to be $\sim$20\% larger than the 
direct measurement.

We have finally carried out the same comparison for Megacam
sensors. We have taken the star size measurements from Tab. 4 of
\cite{Astier13}, and the brighter-fatter slope predicted from
flat-fields correlations is $\sim$30\% smaller than these
measurements. However, the comparison is here limited by the small
amount of flat-field data we have been able to use, and by the fact
the brighter-fatter effect is small (less than 0.5\% over the full
scale) and somehow ``hidden'' by more trivial variations of the
object's sizes with color, coupled to flux-color trends in the star
sample.

\section{Discussion}
\label{sec:discussion}
Coulomb forces within CCDs are a plausible cause of the
brighter-fatter effect and statistical correlations observed in flat-fields.
The latter are also related to the flattening of the PTC 
commonly observed with CCD devices. We propose a scheme to describe
the brighter-fatter effect at the pixel level using measured correlations in
flat-fields that matches measurements at a typically $\sim$20\% accuracy.

Further work is required before we can practically handle the
brighter-fatter effect along this proposed line. We have neglected the
fact that charges in the CCD not only alter drift lines but also
reduce the longitudinal drift field (this is usually referred to as
``space charge effect'').  As a result, the measured correlations in
flat-fields are slightly increased by the effect, generally absent in
science data because the sky level is most often deliberately kept low
by adjusting the exposure time. This decrease of the longitudinal
drift field also slightly increases the lateral diffusion for bright
spots with respect to faint ones (see the contribution by S. Holland
at this workshop). However, the size of this effect is much smaller
than what we observe, in particular for the practical case of a PSF
broader than 3 pixels FWHM. Among the three sensor types we have
discussed, the one from DECam is the most affected by these
space-charge effects because it has the lowest nominal drift field:
the bias voltage is 40 V for a thickness of 250~$\mu$m.

Properly handling the brighter-fatter effect aims at evaluating the
PSF shape at all fluxes where it is practically needed. The
non-linearity of response of the sensor (typically due to the
electronic chain) can also produce variations of the PSF size with
flux, and has to be handled separately. In our comparisons, we have
ignored possible non-linearities. For Megacam, there is practical
evidence that they are very small (see Appendix C
in \cite{Regnault09}). For the E2V CCD data, variations of the light
source intensity with time limit our capability to address
linearity. For DECam, a significant non-linearity has been detected
(see the contribution of G. Bernstein at this workshop) by regressing
the average measured level of dome flat-fields against exposure
time. One might wonder if the electronic chain responds in a similar
way to high steady fluxes (as for high-level flat-fields) and to
localized high signal levels in a few pixels (as for bright stars). Taking these
non-linearity measurements at face value increases the observed
non-linearity but the disagreement between measured and simulated
bright-fatter slopes remains of similar size, with the opposite sign.

One obvious weakness of the proposed approach is that the general
model has more coefficients than the number of two-point flat-field
correlations we can measure. One possible way out is to improve
electrostatic simulations in order to derive relations between model
coefficients which are reasonably independent of the unknowns
affecting the sensor layout. Regarding measurements, we have not
settled yet if higher order flat-field correlations deliver new
information compared with two-point correlations, nor if higher order
correlations are practically measurable. One can also consider
measuring the response of sensors to other patterns than just
flat-fields to constrain the evolution of shapes of sharp spots. How
the brighter-fatter effect varies with image quality likely contains 
some amount of information.

Regarding the practical handling of the brighter-fatter effect in
reduction pipelines, we can already imagine two avenues: the simple
one consists in ``de-scrambling'' the images at the pixel level, using
some inverse of relation \ref{eq:full-model}. If just flipping signs
of the $a$ coefficients is not accurate enough, some refinement can
easily be devised. However, this model assumes that the image is
properly sampled, i.e. that the flux on pixel boundaries is well
approximated by interpolating between the pixel contents. This is
likely to be an unacceptable approximation for sharp images as
typically obtained from space. One can then account for the
brighter-fatter effect inside the PSF model, using the PSF itself to
evaluate the charge density along pixel boundaries.

It is likely that the crude technique we have exposed in this work can
handle the brighter-fatter effect at the $\sim$10\% accuracy, with
some refinements, possibly along the lines mentioned in the above
paragraphs. It is certainly premature to consider that a 1\% accuracy
is within reach. In particular, the brighter-fatter effect is
anisotropic, because the pixel boundaries separating line and rows of
CCDs are generated via totally different mechanisms for most CCD
types. We believe that the way to account for the evolution of PSF
ellipticity with flux at the $\sim 10^{-4}$ level, as required for
very large scale weak shear programs (see \cite{EuclidRB}), is still
an open question.

\acknowledgments
We thank the organizers of the workshop for inviting us to
participate.  Before and after the workshop, we had fruitful exchanges
with R. Lupton.  We used freely available science verification DECam
data.  We had several useful discussion with members of the DES
collaboration, in particular D. Tucker and G. Bernstein who are in
close contact with the data, and J. Estrada for his deep knowledge of
the DECam sensors. DECam non-linearity of response was derived by H. Lin.
We used Megacam science data from the CFHTLS, now freely available.
The Megacam flat-field raw was made available to us by
J.-C. Cuillandre.

\def\aap{{\em A\&A}}
\def\apj{ApJ}
\def\apjl{ApJ Lett.}
\def\apjs{ApJ Supp.}
\def\aj{AJ}
\def\prd{Phys. Rev. D}

\bibliographystyle{jhep}
\bibliography{biblio}

\end{document}